\documentclass[referee,sn-apa]{sn-jnl}

\usepackage{graphicx}%
\usepackage{multirow}%
\usepackage{amsmath,amssymb,amsfonts}%
\usepackage{amsthm}%
\usepackage{mathrsfs}%
\usepackage[title]{appendix}%
\usepackage{xcolor}%
\usepackage{textcomp}%
\usepackage{manyfoot}%
\usepackage{booktabs}%
\usepackage{algorithm}%
\usepackage{algorithmicx}%
\usepackage{algpseudocode}%
\usepackage{listings}%
\usepackage{tabularx}%
\usepackage{amsmath, amssymb, bm, bbm}%
\usepackage[12pt]{extsizes} 

\begin{document}

\title[Article Title]{Variable selection via knockoffs in missing data settings with categorical predictors}

\author*[1]{\fnm{Silvia} \sur{Bacci}}\email{silvia.bacci@unifi.it}

\author[1]{\fnm{Emanuela} \sur{Dreassi}}\email{emanuela.dreassi@unifi.it}

\author[1]{\fnm{Leonardo} \sur{Grilli}}\email{leonardo.grilli@unifi.it}

\author[1]{\fnm{Carla} \sur{Rampichini}}\email{carla.rampichini@unifi.it}

\affil*[1]{\orgdiv{Department of Statistics, Computer science, Applications}, \orgname{University of Florence}, \orgaddress{\street{Viale Morgagni, 59}, \city{Florence}, \postcode{I-50134}, \country{Italy}}}

\def\baselinestretch{2.0}\def\arraystretch{0.7}

\abstract{
Large-scale assessment data typically include numerous categorical variables, often affected by missing values. Motivated by the challenges arising in this framework, we extend the knockoffs method for selecting predictors to settings with missing values. 
Our proposal relies on a preliminary phase consisting of multiple imputations of missing values. Each imputed dataset is then processed using a suitable knockoff filter.  
We evaluate the performance of the proposed method through a simulation study, showing satisfactory results consistent with 
a recently advocated cutting-edge method.
We apply the method to large-scale assessment data collected by INVALSI about test scores of Italian students in grade 5 with many background variables. This case study is challenging, as most predictors have unordered categories, a setting not taken into account by traditional knockoffs methods. In addition, some of the key predictors are affected by missing values. The model includes random effects to account for the multilevel structure of students nested into schools.
Our proposal to implement the knockoffs method within a multiple imputation framework proves to be feasible, flexible and effective. 
}


\maketitle

\section{Introduction}
\label{sec:intro}

The selection of predictors with a relevant effect on an outcome is a fundamental issue in assessing a statistical model. It is particularly challenging when numerous predictors are available. Indeed, different selection strategies may lead to different results with the risk of including in the model \textit{null variables}, i.e., predictors with a regression coefficient equal to 0 in the true model or, on the opposite, excluding \textit{non-null variables}.

This issue is encountered in many contexts, including education, where large-scale assessment data are routinely collected to measure students' proficiency levels. In Italy, the National Institute for the Evaluation of the Education and Training System (INVALSI, \url{https://www.invalsiopen.it/}) administers standardized tests to assess ability in Italian, Mathematics, and English to students of different grades, starting from primary school to the end of high school. 
INVALSI collects many variables about the students and their families. Some of the variables, such as parents' education, have been widely recognized as determinants of student achievement, whereas other variables are just proxies and they should be inserted in the model only if predictive of the outcome (e.g. the number of books at home, or the availability of a room for studying). It is, therefore, essential to exploit effective variable selection methods. The challenge is to apply those methods in this setting where most of the variables are affected by missing values, with rates as high as about 26\%.

In the literature, many methods for variable selection in regression models have been proposed, starting from the classical forward, backward and stepwise selection, to the modern regularization approaches such as the lasso \citep{tibshirani:1996} and the knockoff filter. The knockoffs method \citep{barber:candes:2015} 
is based on the general idea of randomly generating variables (i.e., knockoffs) that mimic the original predictors in terms of distributive properties but 
are conditionally independent of the response variable given the original variables. 
Then, the importance of these copies is compared to that of the original variables through a suitable statistic based on the comparison of the related regression coefficients. 
As any other selection method, the knockoff filter aims at identifying the relevant (i.e., {\em non-null}) variables, but it has the peculiarity of controlling for the False Discovery Rate (FDR), that is 
the expected proportion of false discoveries (i.e., null variables wrongly declared as non-null), or, alternatively, 
the expected number of false discoveries, that is the Per Family Error Rate (PFER). We stress that the control of FDR (or PFER) is particularly important in high-dimensional settings, where the large number of available predictors implies a severe risk of selecting non-relevant (i.e., {\em null}) variables. 
 
Numerous contributions have followed within a few years in the framework of the knockoff literature. 
However, traditional knockoffs methods are limited to continuous predictors and complete data, thus they are not fully appropriate in many real-world applications. A recent extension to treat mixed types of variables is due to \cite{kormaksson:2021}, which, however, still requires complete data. To our knowledge, the only approach handling both categorical predictors and missing values is \cite{xie:2023}.   

In particular, 
missing values depend only on non-null variables, according to the assumption of Strong Missing At Random (SMAR), which is more restrictive than the standard Missing At Random \citep[MAR; ][]{little:rubin:2002}.
This approach relies on a latent variable model, where the predictors are assumed to follow a multivariate normal distribution, which is then categorized to handle binary and ordinal observed variables. The latent variable model is estimated through maximum likelihood, and the missing values are integrated out via an Expectation-Maximization (EM) algorithm \citep{demp:lair:rubi:77}. In summary, even if \cite{xie:2023} specify a complex model, their approach is based on restrictive assumptions on both the missingness mechanism and on the distribution of the predictors. Moreover, their approach can only handle continuous and binary/ordered predictors, which is a limitation in settings with unordered categorical predictors, such as INVALSI data. 

To overcome the mentioned limitations, we propose to embed the knockoffs method into a Multiple Imputation \citep[MI; ][]{rubin:1987, rubin:1996} 
framework, separating the phase of treatment of missing data from the phase of variable selection. MI is a general and well-established approach to handling missing values under a MAR mechanism.  Since MI produces a number of complete datasets, it is straightforward to apply any standard method to each dataset. The challenge is to choose how to combine the results obtained in the imputed datasets. In fact, for inference on model parameters, the results can be combined with Rubin's rules, ensuring well-defined properties. However, for variable selection methods, there are no theoretical results on the properties of the combined results \citep[see ][and the references therein]{grilli:2022}. Therefore, the choice should be guided by evidence from simulation studies. 

The rest of the paper is organized as follows. 
Section \ref{sec:method} illustrates the proposed knockoffs method with multiple imputation. Section \ref{sec:simulation} summarizes the results of a simulation study to assess the performance of the new method compared with the approach of \cite{xie:2023}. Section \ref{sec:analysis} describes the INVALSI large-scale assessment data and reports the results of the analysis based on the proposed method. Section \ref{sec:final} offers some final remarks.

\section{Variable selection via knockoffs in presence of missing values}
\label{sec:method}
In this section we  review the fundamentals of variable selection using the knockoff filter, starting with the original approach designed for continuous variables, then outlying a sequential knockoffs method which properly handles also categorical variables. Then we illustrate our proposal to exploit multiple imputation to implement the knockoffs filter in presence of missing values.

\subsection{The knockoff filter: methodological background}
\label{sec:methodback}
Among the methods to select the set of {\em non-null} variables, namely the predictors with an effect on the outcome in the population model, the knockoff filter has a very recent history. Its introduction dates back to the work of \cite{barber:candes:2015}, then extended by \cite{candes:2018}. With respect to traditional methods for variable selection (e.g., forward selection, backward elimination), the knockoff filter explicitly controls the PFER or the FDR. Let $S_D$ be the set of discoveries (i.e., selected variables) and $S_F$ be the set  of false discoveries (i.e., selected variables with a null effect), then 
$$
PFER = E(|S_F|)
$$
and 
$$
FDR = E\left(\frac{|S_F|}{|S_D|}\right).
$$

Given a regression model for the outcome $Y$ on the observed $p$ predictors $\mathbf X=[X_1, \ldots, X_p]$, the knockoffs $\tilde{\mathbf X}=[\tilde{X}_1, \ldots, \tilde{X}_p]$ are randomly generated variables independent from $Y$ conditionally on $\mathbf X$ and with the following swapping property, holding for any subset $S \subseteq \{1, \ldots, p\}$:

\begin{equation}
 \label{eq:swap}
 [\mathbf X, \tilde{\mathbf X}]_{swap(S)} \overset{d}{=} [\mathbf X, \tilde{\mathbf X}]
 \end{equation}
where $\overset{d}{=}$ denotes the equality in distribution. The vector  $[\mathbf X, \tilde{\mathbf X}]_{swap(S)}$ is obtained by swapping the entries $X_j$ and $\tilde{X}_j$ for any $j \in S \subseteq \{1, \ldots, p\}$, i.e. the distribution of the augmented vector $[\mathbf X, \tilde{\mathbf X}]$ does not change if the original variables are swapped with the corresponding knockoff copies.

To carry out variable selection, the outcome $Y$  is regressed on the augmented vector of predictors $[\mathbf X,  \tilde{\mathbf X}]$ using a regularization procedure such as the lasso. Then, for each element $X_j$ in $\mathbf X$, a statistic $W_j$ is computed, satisfying the flip-sign property;  see \cite{barber:candes:2015} for details.   
A commonly used statistic $W_j$ with the flip-sign property
is the difference between the absolute values of the estimated regression coefficients $\hat{\beta}_j$ and $\hat{\tilde{\beta}}_j$ for $X_j$ and its copy $\tilde{X}_j$, respectively: 
 \begin{equation}
 \label{eq:statw}
 W_j = |\hat{\beta}_j| - |\hat{\tilde{\beta}}_j|.
 \end{equation}
The larger the value of $W_j$, the stronger the evidence in favor of the relevance of $ X_j$. Specifically, the variable $X_j$ is selected if $W_j >t$, where $t$ is a constant chosen to ensure that the PFER or the FDR is not greater than the desired level. 

Since the set of selected variables according to the original knockoffs approach may depend on the specific values randomly generated for the knockoffs $\tilde{\mathbf X}$, in the derandomized knockoffs approach \citep{ren:2023} the non-uniqueness of the knockoff solution is addressed repeating the described procedure a certain number of times (usually, $31$). Then, variables chosen in at least a given proportion of cases (usually, $0.50$) are selected. 

The cited knockoff methods implicitly assume continuous predictors. To allow for a wider applicability of the knockoffs filter approach,  \cite{kormaksson:2021} proposed an algorithm to treat both continuous and categorical variables. The proposed sequential knockoff method differs in the procedure used to generate knockoffs. Specifically, the knockoff $\tilde{X}_j$ is randomly sampled from a normal distribution when $X_j$ is a continuous variable and from a multinomial distribution when $X_j$ is a categorical variable. The characteristic parameters of these distributions (i.e., mean and variance in the normal case and vector of probabilities 
in the multinomial case) are estimated through a sequential procedure: for $j=1, \ldots, p$, the predictor  $X_j$ is regressed on the remaining predictors $\mathbf{X}_{-j}$ and on the knockoffs generated in the previous steps $[\tilde{X}_1, \ldots, \tilde{X}_{j-1}]$. The model for the knockoff generation is a penalized linear model if $X_j$ is continuous and a penalized multinomial logit model if $X_j$ is categorical.

After the generation of knockoffs, the comparison among regression coefficients $\hat{\beta}_j$ and $\hat{\tilde{\beta}}_j$ proceeds along the same lines as \cite{candes:2018}, with multiple knockoffs generation and, consequently, multiple selections as in \cite{ren:2023}.  

It is worth to note that when a categorical variable with $K$ levels is used as a predictor, it enters the model as a set of $K-1$ dummy variables. The standard lasso treats those dummy variables as distinct predictors, so it may happen that only some of them are selected, depending on the arbitrary choice of the reference category. To avoid this issue, one can rely on a grouped lasso \citep{yuan2006} that treats the set of dummies as a whole. Under grouped lasso, the statistic $W_j$ in \eqref{eq:statw} modifies as follows:
\begin{equation}
\label{eq:statw2}
 W_j = \max_k\left\{\left||\hat{\beta}_{jk}| - |\hat{\tilde{\beta}}_{jk}|\right|\right\},
\end{equation}
where $\hat{\beta}_{jk}$ and $\hat{\tilde{\beta}}_{jk}$ are the  estimated regression coefficients of the dummy for category $k$ ($k = 2, \ldots, K$) of variable $j$ and its knockoff, respectively.
Other approaches to deal with grouped variables in the knockoffs framework are described in \cite{dai:barber:2016} and \cite{katsevich:sabatti:2019}, though they explicitly consider only  continuous variables. 

The sequential knockoffs procedure was recently optimized from a computational perspective by \cite{zimmermann:2024} through the development of the sparse sequential knockoff algorithm, which introduces a preliminary phase aimed at pre-selecting predictors. This phase relies on the graphical lasso \citep{friedman:2008} to estimate the precision matrix of the original set of predictors. Zero entries in the precision matrix are used to identify predictors that are conditionally independent of each variable $X_j$, and these predictors do not enter in the subsequent step of generation of $\tilde{X}_j$. Since the graphical lasso is designed for continuous variables, \cite{zimmermann:2024} address the issue of categorical variables by employing a dummy encoding scheme. In this approach, conditional dependence between two categorical variables is assumed if at least one non-zero element appears in the precision matrix of the dummy-encoded representation. 

All the methods described so far require complete data sets. To overcome this limitation, \cite{xie:2023} proposed a general setting that extends the derandomized approach of \cite{ren:2023}, allowing to accommodate both missing values and observed variables of mixed type (continuous, binary and ordered).   Their proposal relies on a restrictive version of the MAR assumption \citep{little:rubin:2002}, 
known as Stronger MAR (SMAR), according to which the probability of missingness depends only on non-null variables. To accommodate in the same model continuous, binary, and ordinal variables, the approach is based on a latent variable model where the predictors are assumed to be jointly distributed as an underlying multivariate Normal. This model is estimated through the maximum likelihood approach by integrating out the missing values through an EM algorithm \citep{demp:lair:rubi:77}.  Currently, the approach does not allow for unordered categorical and count variables or less restrictive assumptions about missingness. This approach which simultaneously specify a model for the outcome and integrates out the missing data via the EM algorithm is very efficient if the underlying assumptions are satisfied, but is difficult to extend to different settings.

\subsection{Implementing the knockoff filter with multiple imputation}
\label{sec:methodproposal}
To extend the range of possible models and enable flexible handling of missing data, we propose a multi-step procedure for variable selection using knockoffs, based on the following steps:
\begin{description}
\item[(i)] MI of the missing data; 
\item[(ii)] application of the knockoff filter to each imputed dataset;
\item[(iii)] identification of a unified set of selected variables.
\end{description}
Once the variable selection process is concluded, the model of interest is fitted on the imputed datasets using the set of selected variables.

The first step of the proposed procedure consists of imputing missing values through MI \citep{rubin:1987, rubin:1996}, a general and well-established approach to handling missing values under a MAR mechanism.
The implementation of the MI procedure involves several critical choices. Among these, the selection of the imputation method and the number of imputed datasets are particularly important. We adopt Multiple Imputation by Chained Equations \citep[MICE; ][]{buuren:2011, buuren:2018}, a widely used and flexible approach that accommodates various types of variables and complex relationships among them. MICE is particularly advantageous in settings where data include both categorical and continuous variables, as it imputes missing values sequentially for each variable conditional on all others, using tailored imputation models. This flexibility makes MICE suitable for INVALSI data, characterized by a set of  ordered and unordered categorical variables with varying degrees of missingness.

Another critical decision concerns the number of imputed datasets. A small number of imputed datasets, say 5, is often enough to appropriately account for the uncertainty due to the imputation process.   However, the optimal number of imputed datasets depends on various factors, including the proportion of missing data, the mechanism underlying the missingness, and the desired precision of estimates \citep{graham:2007}.
The search for the optimal number of imputed datasets is beyond the scope of this paper. In the simulation study of Section \ref{sec:simulation} and in the application of Section \ref{sec:analysis} we use 10 imputed datasets,  to get a balance between computational burden and efficiency \citep{white:2011, boden:2011}.

After imputing the missing values using MI, variable selection is performed separately for each imputed dataset using the knockoff approach. It is important to note that any  knockoff method can be applied. In the  simulation study of Section \ref{sec:simulation}, we compare the performance of the proposed MI based procedure using two types of knockoffs: (i) the knockoffs procedure of \cite{ren:2023} that is the one embedded within the algorithm of \cite{xie:2023}, enabling a 
comparison with our proposed method, under the same conditions, and (ii) the sequential knockoffs procedure of  \cite{kormaksson:2021} and 
\cite{zimmermann:2024} 
that properly handles  categorical variables, making it particularly suitable for application to INVALSI data.

The knockoffs based variable selection is applied to each imputed dataset separately. 
In this way, the FDR is controlled within each imputed dataset, but there is no guarantee to control the {\em overall} FDR.
 Each variable can be always selected, never selected or selected in a proportion of the imputed datasets. Thus, the third step involves choosing the {\em selection proportion}, namely the minimal proportion for a variable to be ultimately selected. 
A higher selection proportion reduces the probability of a variable being ultimately selected, which in turn decreases the expected number of false discoveries (i.e., non-null variables wrongly selected) and, thus, the overall FDR. Thus, a conservative choice would be to set the minimal proportion to 1, i.e. ultimately select only variables selected in all the imputed datasets. However, this choice could lead to a final FDR much lower than the target one at the cost of a reduction of the ability of the procedure to correctly identify non-null variables, leading to a potentially low True Positive Rate (TPR). 

In the context of variable selection via stepwise,  \cite{wood2008} suggest ``the majority rule'', that is, the selection of a variable whenever it is chosen in the majority of the imputed datasets. However, their study does not investigate the impact of this strategy on the overall FDR.
As there are no further theoretical guidelines for determining the optimal selection proportion, 
in the simulation study presented in the next section, we examine the results for different values of the selection proportion.

\section{Simulation study}
\label{sec:simulation}

To investigate the properties of our proposal to embed the knockoffs method into an MI procedure, we perform a  Monte Carlo simulation study. The main features of the simulation study are reported in Table \ref{tab:simudesign}.

\begin{table}[!ht]
\caption{Overview of the simulation study}
\label{tab:simudesign}
\centering
\begin{tabularx}{\textwidth}{l X}
\toprule
{Methods}   
        & - XCDW: Maximum Likelihood with EM, missing values integrated out \citep{xie:2023} - \texttt{R} scripts made available by \cite{xie:2023} \\
        & - MI-lasso: MICE + grouped lasso \citep{yuan2006} - \texttt{R} packages \texttt{mice} \citep{buuren:2011} and \texttt{grpreg} \citep{breheny:2015} \\
        & - MI-RWC: MICE + derandomized knockoffs \citep{ren:2023} - \texttt{R} packages \texttt{mice} and \texttt{derandomKnock} \citep{ren:2023} \\
        & - MI-seq: MICE + sparse sequential knockoffs \citep{zimmermann:2024} - \texttt{R} packages \texttt{mice} and \texttt{knockofftools} \citep{zimmermann:2024} \\
\midrule
        {Tuning parameters} & - PFER = 2, except for MI-lasso which is based on minimum BIC \\
            & - Multiple runs (except for MI-lasso): 31 runs with selection threshold    0.5
            \\
\midrule
{Sample size} & 1000 observations \\
\midrule
{Variables} & 100 (50 binary, 50 continuous) divided into 5 blocks of 20 variables with the correlation structure of \cite{xie:2023} \\
 & - Non-null variables:   2 for each block (1 binary, 1 continuous), for a total of 10 (5 binary, 5 continuous)\\
\midrule
        {Missing data mechanism} & - SMAR: missingness only depending on non-null observed variables (1 binary and 1 continuous for each individual) \\
                    & - MAR: missingness depending on both non-null and null observed variables (1 binary non-null, 1 continuous non-null, 1 binary null, and 1 continuous null, for each individual) \\
\midrule
 Rate of missing values &   0\% (complete data), 10\%, 32\%, 45\%  \\
 \midrule
 {Number of imputed datasets} &    10 (except for XCDW) \\
 \midrule
{Monte Carlo} & 100 runs \\
\bottomrule
\end{tabularx}
\end{table}

We implement the MI-based knockoff approach using both the derandomized knockoff filter proposed by \cite{ren:2023} (referred to as MI-RWC) and the sparse sequential knockoff filter introduced by 
 \cite{zimmermann:2024} (referred to as MI-seq).
It is worth to outline that in both cases the selection procedure is nested: first, the multiple repetition of the knockoff process (31 runs) selects the variables {\em within} each of the 10 imputed datasets; then, these selections  are combined across the imputed datasets to obtain a unique set of selected variables.
 
The results obtained by these MI-based methods (MI-RWC, MI-seq, and MI-lasso) are compared with those obtained using the approach of \cite{xie:2023}, here referred to as XCDW, which is currently the only well-established method for handling knockoffs with missing data.   To this end, we devise a simulation study closest to the one of \cite{xie:2023}, starting from their scripts. 

To make  the three approaches comparable, we set the nominal value of PFER (i.e., 2), which is the only metric that can be controlled by any of the three approaches. For details on how PFER is controlled by the algorithms, see the specific references \citep{ren:2023, xie:2023, kormaksson:2021}.

In the simulation study, 
we also apply a standard selection procedure based on lasso (without a knockoff filter) to the imputed datasets, denoted as MI-lasso. 

The simulated scenario consists of 50 binary variables, 5 of which are non-null, and 50 continuous variables, 5 of which are non-null, with the correlation structure specified 
as in \cite{xie:2023}. In detail, the 100 variables are equally divided into five blocks (each block contains 10 binary variables and 10 continuous variables). The correlation within each block is: 0.60 for the first, second, and third blocks, and 0.30 for the fourth and fifth blocks; the correlation between pairs of blocks ranges from 0.10 to 0.30. For further details, see \cite{xie:2023} (Section 5, Figure 4). 
The response variable is generated by a linear regression model with a normal error and with coefficients of non-null variables equal to $+0.5$ for binary ones and $-0.5$ for continuous ones.

Missing values on predictors are generated using the SMAR mechanism \citep[as described in][]{xie:2023}, where missingness depends solely on non-null observed variables associated with the outcome. Specifically, in the \cite{xie:2023}'s simulation scheme, for each individual $i$ two of the ten non-null variables, say $X_{ij_1}$ and $X_{ij_2}$, are assumed to be observed and affect  probability of  missing value of the remaining 98 ones, as follows
\begin{equation}
P(X_{ij} = NA) = \frac{1}{1+ e^{-\left[h + \left(X_{ij_1} + X_{ij_2})/2\right) \right]}}, \quad j \neq j_1, j_2,
\label{eq:miss}
\end{equation}
with  $h$  constant that drives the rate of missing values. 

Additionally, since MI is valid under the more general MAR assumption, we introduce a MAR scenario by allowing missingness on predictors to depend also on two null observed variables, in addition to the two non-null ones. Formula \eqref{eq:miss} modifies accordingly by adding two further addends and dividing by 4.

Setting $h = -1$ in formula \eqref{eq:miss} as in \cite{xie:2023},
the resulting rates of missing values in the simulations are approximately 32\%. 
In the following we present the results in detail for this scenario (Section \ref{sec:ressim1}). Then, we extend the simulation  using also different values of constant $h$ to change the rate of missingness (Section \ref{sec:ressim2}). Specifically, we adopt $h = -2.4, -0.4$ that result in about  10\% and 45\%  of missing values, respectively. 

The simulation results are summarized by the following indexes, where `discoveries' stands for `selected variables' and MC for Monte Carlo mean:
\begin{eqnarray*}
 \widehat{PFER} & = &  MC(\textrm{\# false discoveries}) \\ 
\widehat{FDR}  & = & MC\left(\frac{\textrm{\#  false discoveries}}{ \textrm{\# discoveries}}\right) \\
\widehat{TPR} & = & MC\left(\frac{\textrm{\#  true discoveries}}{\textrm{\#  non-null variables}}\right),
 \end{eqnarray*}
Note that the three indexes $\widehat{PFER}$, $\widehat{FDR}$, and $\widehat{TPR}$ represent the empirical counterparts of the theoretical ones.

\subsection{Results of Monte Carlo simulations for 32\% of missing values} \label{sec:ressim1}

Table \ref{tab:simuresults} reports  
the values of  $\widehat{PFER}$, $\widehat{FDR}$, and $\widehat{TPR}$ and their standard deviations for the methods under comparison.
For MI-based approaches, the performances are evaluated considering different  
selection proportions over the 10 imputed datasets. First, we comment on the PFER, which is the tuning parameter of the knockoff filters used here; then we consider the FDR and TPR. 

The XCDW method has an empirical PFER of 2.12 under SMAR and 2.51 under MAR, close enough to the nominal value of 2. On the other hand, for MI-based approaches, the empirical PFER depends on the selection proportion over the imputed datasets: specifically, for MI-RWC the empirical PFER gets close to 2 when the proportion is one, 
whereas for MI-seq the optimal proportion is between 0.8 and 0.9. In the MI-lasso the tuning parameter is chosen minimizing BIC, thus it does not control the PFER; in the simulation study, MI-lasso turns out to have values much greater than 2 even when the selection proportion is one.

\begin{figure}[!ht]
  \centering
   \begin{tabular}{cc}\vspace{-0.5cm}
  \includegraphics[scale=0.25]{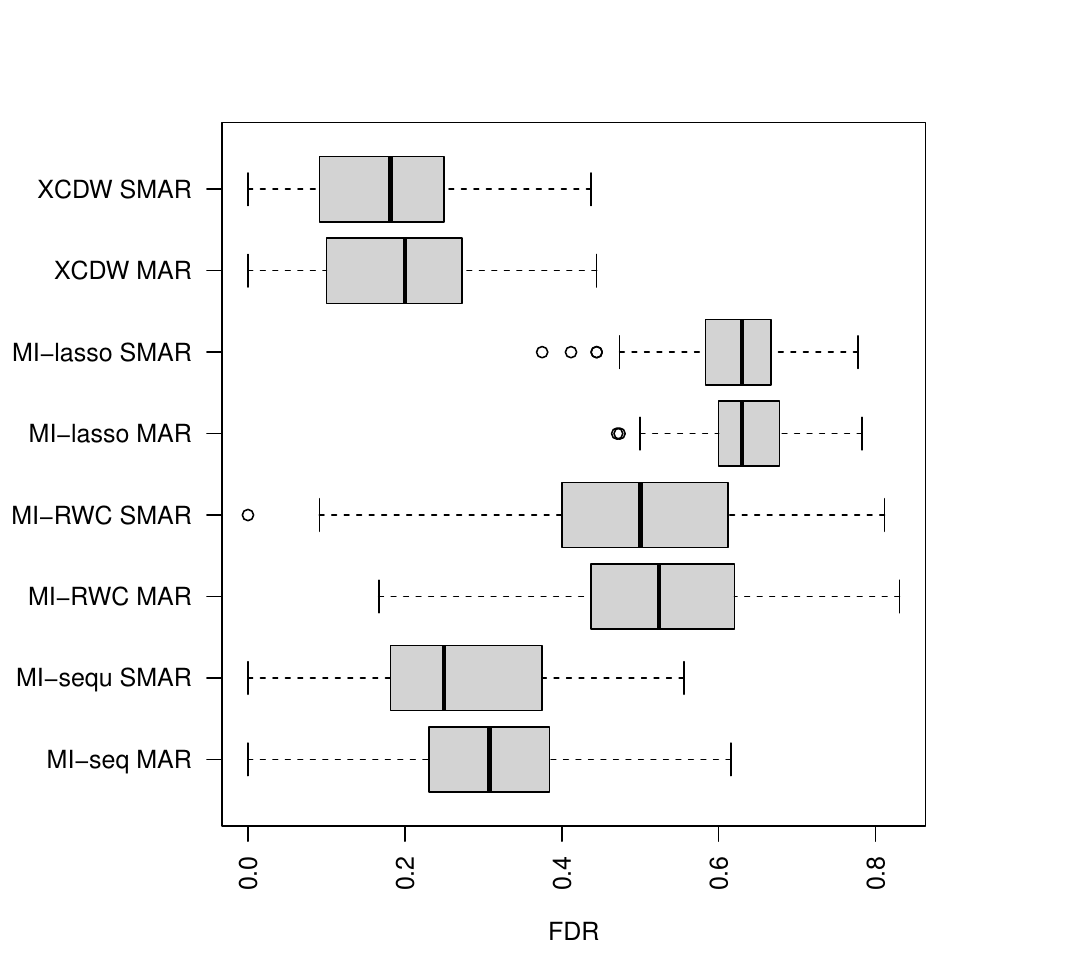} &\includegraphics[scale=0.25]{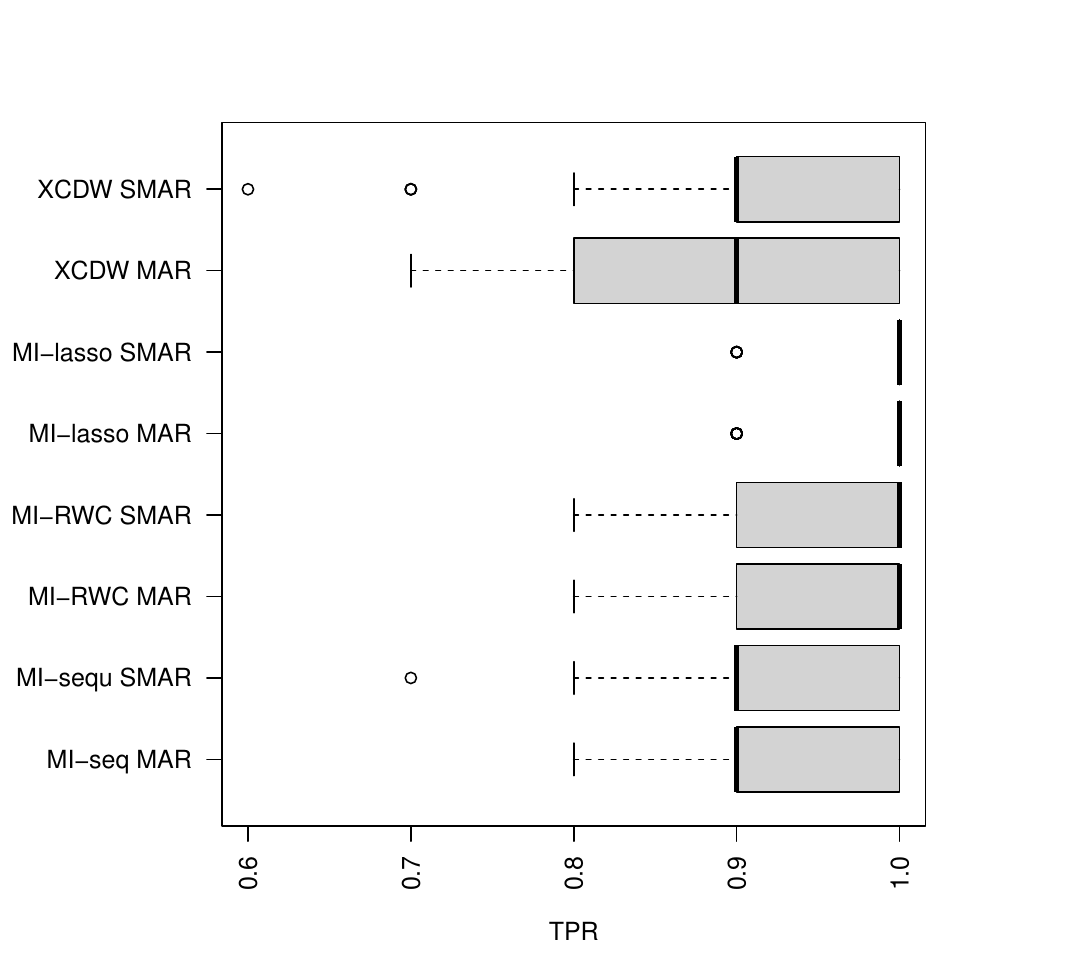} $\geq 70\%$\\
  \vspace{-0.0cm}
   \includegraphics[scale=0.25]{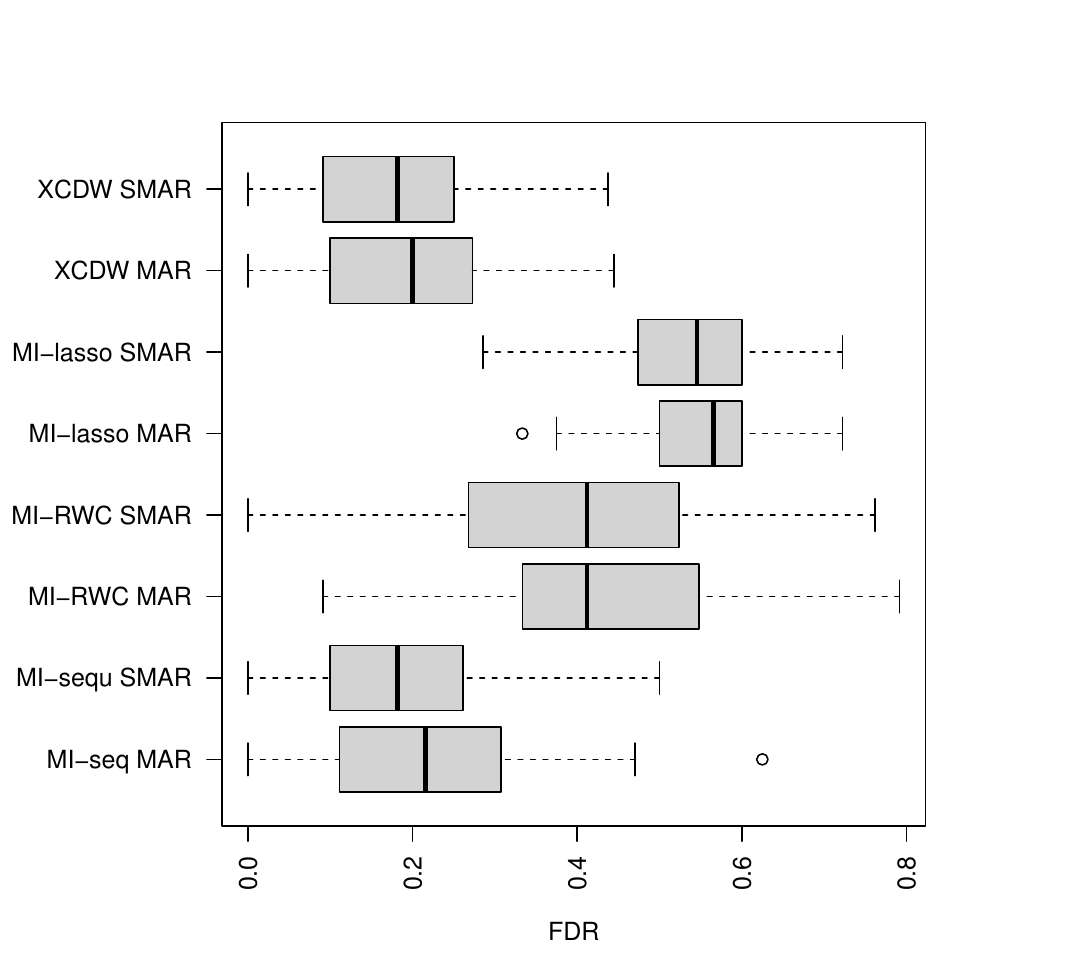} &
   \includegraphics[scale=0.25]{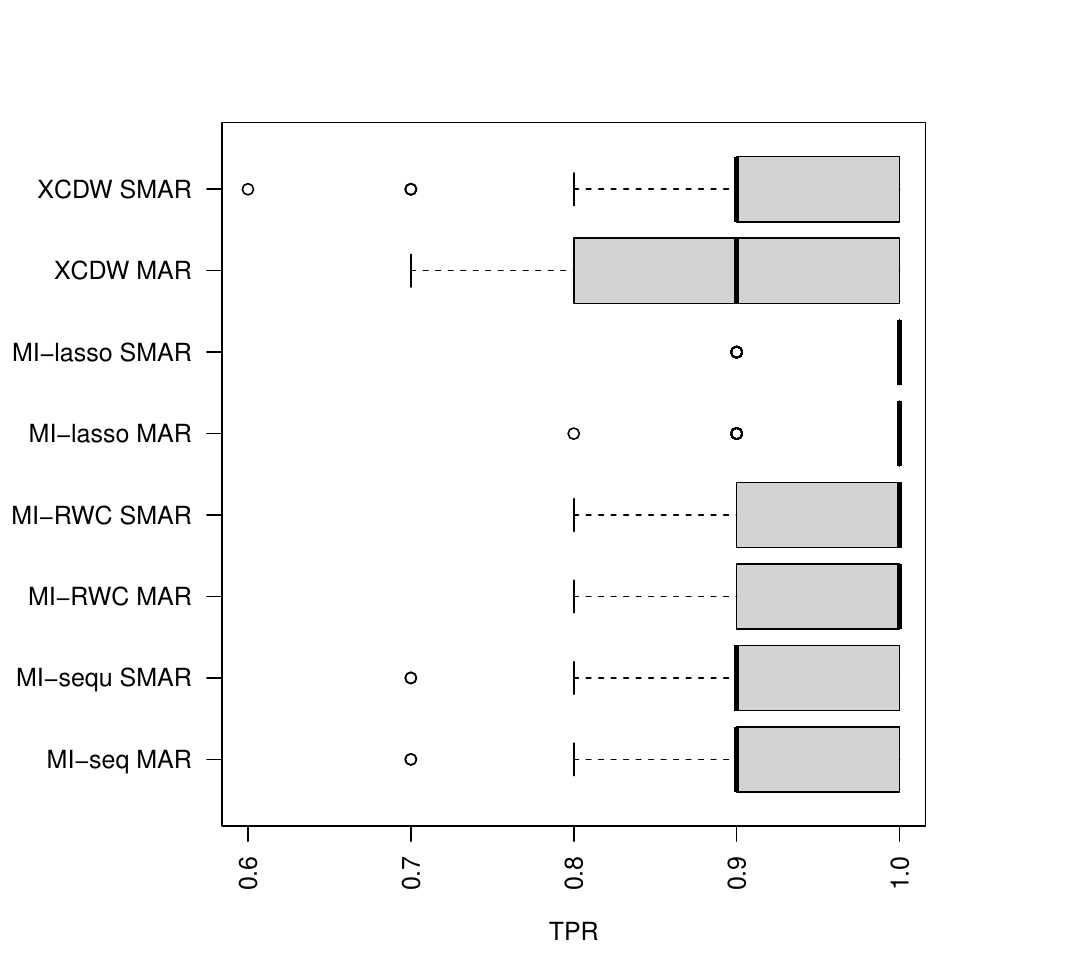}$\geq 80\%$ \\
   \vspace{-0.0cm}
    \includegraphics[scale=0.25]{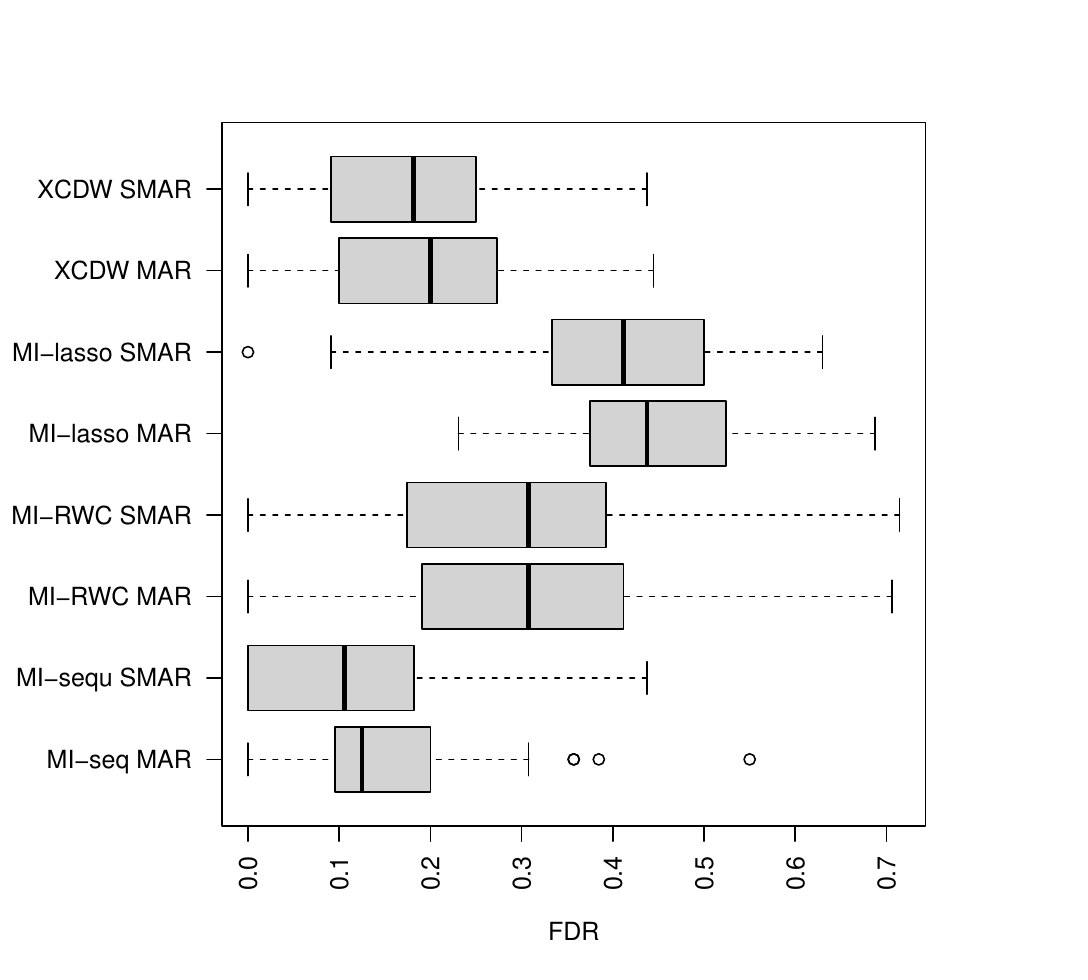} &
    \includegraphics[scale=0.25]{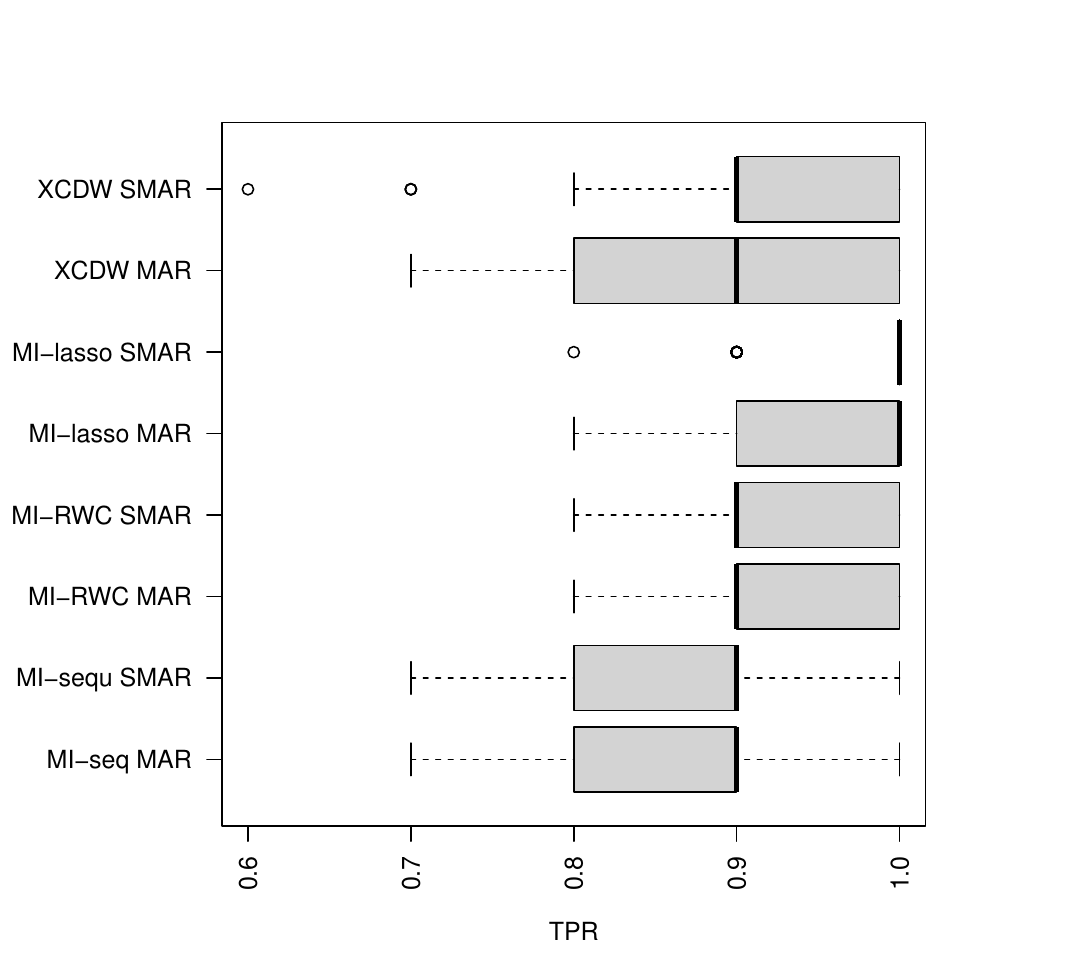}$\geq 90\%$ \\
    \vspace{0.2cm}
     \includegraphics[scale=0.25]{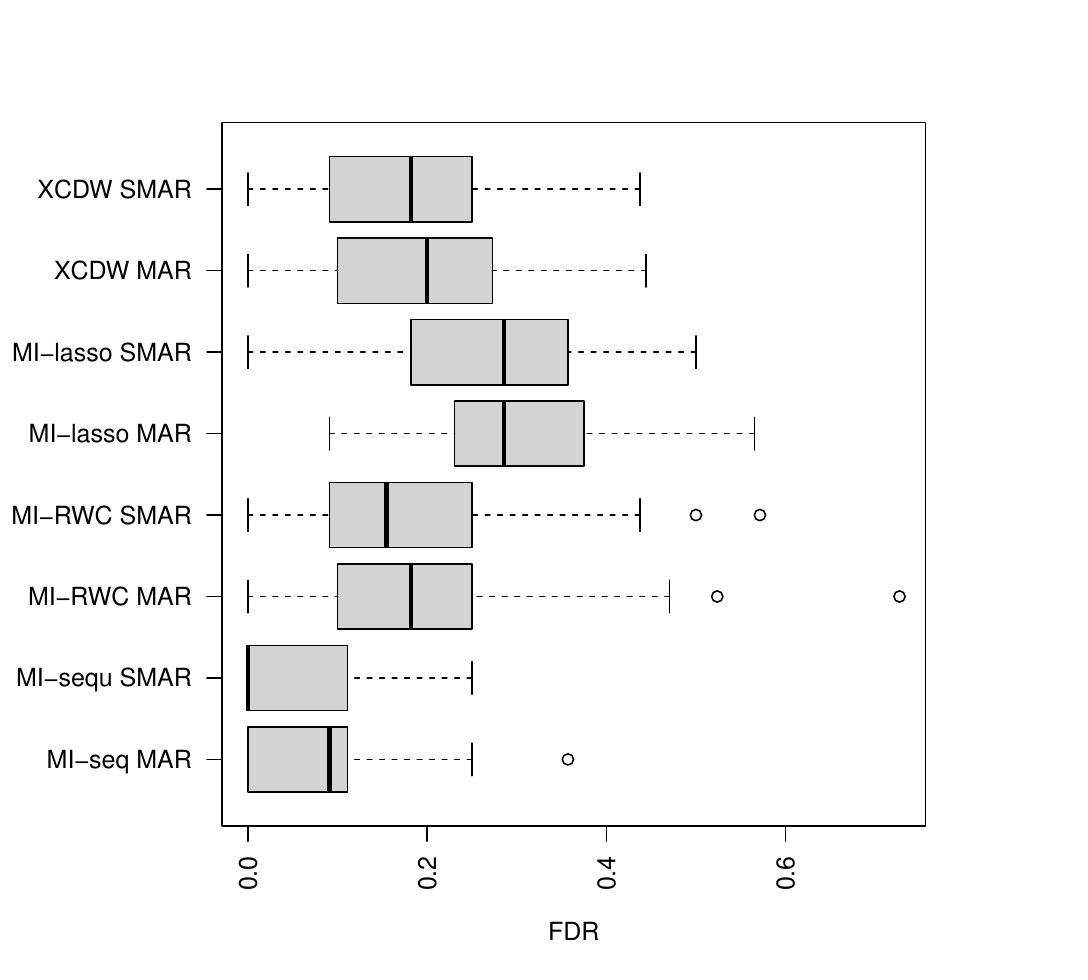} &
     \includegraphics[scale=0.25]{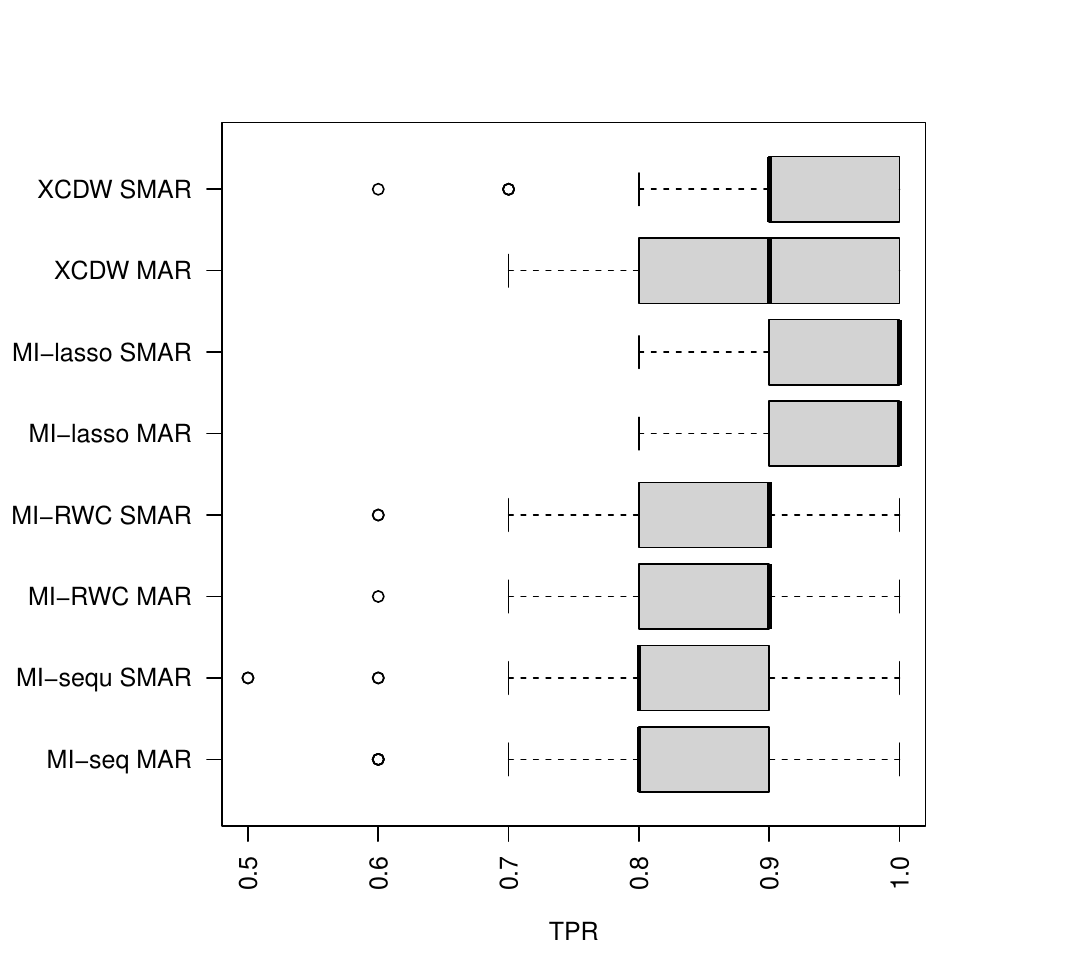}$= 100\%$ \\
      \end{tabular}
  \caption{Distribution of FDR (left) and TPR (right) under XCDW, MI-lasso, MI-RWC, and MI-seq, for an increasing selection proportion 
  (up to bottom panels:  $\geq 0.7$, $\geq 0.8$, $\geq 0.9$, and $1$): box-plots relied on Monte Carlo values for minimum, 1st quartile, median, 3rd quartile, maximum.}\label{fig:boxplot}
\end{figure}

\begin{figure}[!ht]
  \centering
  \begin{tabular}{cc}\vspace{-0.5cm}
   \includegraphics[scale=0.35]{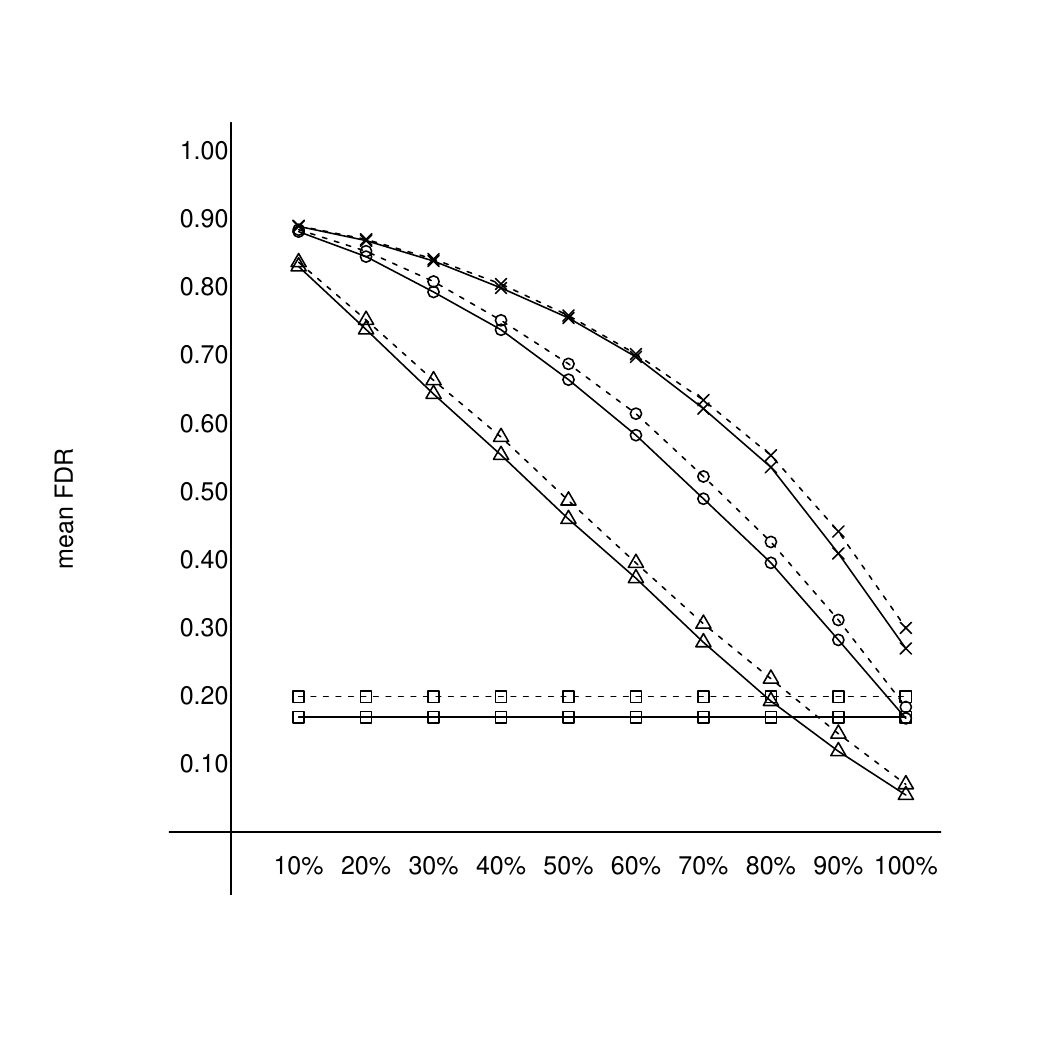} & \includegraphics[scale=0.35]{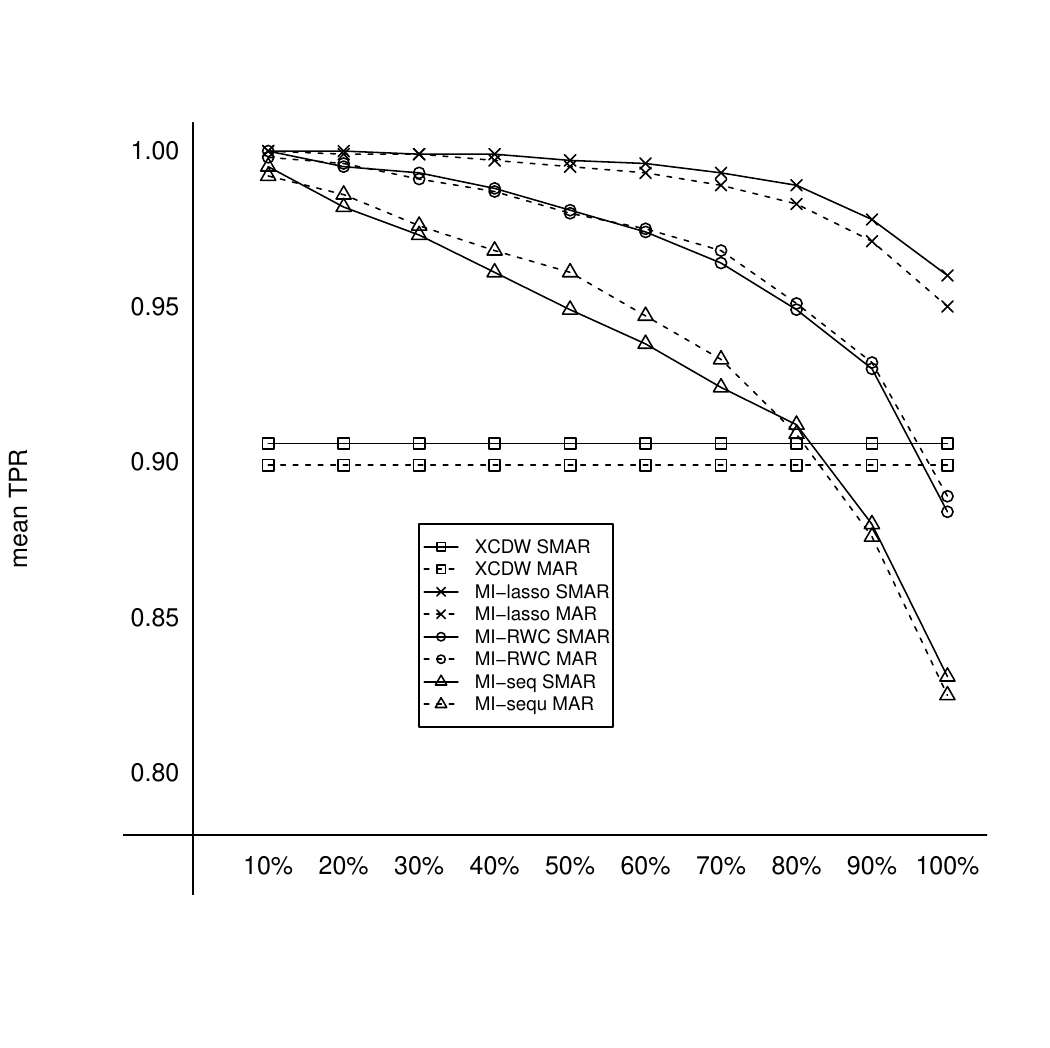} \\
    \footnotesize{FDR} & \footnotesize{TPR}\\
    \end{tabular}
  \caption{Variable selection with XCDW, MI-lasso, MI-RWC, and MI-seq, for an increasing selection proportion for ultimately selecting the variables over the imputed datasets: Monte Carlo means of FDR (left) and TPR (right).}\label{fig:cut}
\end{figure}

Figure \ref{fig:boxplot} displays the box-plots based on the quartiles for FDR and TPR, while Figure \ref{fig:cut} depicts the trends in the mean Monte Carlo values for the two indices.

\begin{table}[!ht]
\caption{Variable selection with XCDW, MI-lasso, MI-RWC, and MI-seq, for 
increasing selection proportion: 
Monte Carlo means and standard deviations of PFER, FDR, and TPR. \textcolor{purple}{Rate of missing values: about 32\%.}} 
\label{tab:simuresults}%
\begin{tabular}{@{}llrrrrrr@{}}
\toprule
Method	&	Missing	&	\multicolumn{2}{c}{PFER}			&	\multicolumn{2}{c}{FDR}	&			\multicolumn{2}{c}{TPR}			\\
	&	mechanism	&	Mean	&	SD	&	Mean	&	SD	&	Mean	&	SD	\\
\midrule															
XCDW	        &	SMAR	 &	2.12	&	1.87	&	0.17	&	0.13	&	0.91	&	0.08	\\
	        &	MAR	     &	2.51	&	1.80	&	0.20	&	0.12	&	0.90	&	0.08	\\
{\em Selection prop.: $\geq 0.6$} 	&		&		&	         	&		&		&		&		\\
$\;\;$MI-lasso	&	SMAR	&	24.04	&	6.15	&	0.70	&	0.05	&	1.00	&	0.02	\\
	          &	MAR	   &	24.37	&	6.34	&	0.70	&	0.05	&	0.99	&	0.03	\\
$\;\;$MI-RWC	&	SMAR	&	16.06	&	8.74	&	0.58	&	0.13	&	0.97	&	0.05	\\
	        &	MAR	&	18.06	&	10.22	&	0.61	&	0.11	&	0.98	&	0.05	\\
$\;\;$MI-seq	&	SMAR	&	6.07	&	2.91	&	0.37	&	0.12	&	0.94	&	0.07	\\
	        &	MAR	&	6.76	&	3.26	&	0.40	&	0.11	&	0.95	&	0.06	\\
{\em Selection prop.: $\geq 0.7$} 	&		&		&	     	&		&		&		&		\\
$\;\;$MI-lasso	&	SMAR	&	17.30	&	5.22	&	0.62	&	0.07	&	0.99	&	0.03	\\
	        &	MAR	&	17.91	&	4.76	&	0.63	&	0.06	&	0.99	&	0.03	\\
$\;\;$MI-RWC	&	SMAR	&	11.20	&	7.09	&	0.49	&	0.16	&	0.96	&	0.05	\\
	        &	MAR	&	12.56	&	7.75	&	0.52	&	0.13	&	0.97	&	0.05	\\
$\;\;$MI-seq	&	SMAR	&	3.93	&	2.35	&	0.28	&	0.12	&	0.92	&	0.07	\\
	        &	MAR	&	4.58	&	2.75	&	0.31	&	0.12	&	0.93	&	0.07	\\
{\em Selection prop.: $\geq 0.8$} 	&		&		&	     	&		&		&		&		\\
$\;\;$MI-lasso	&	SMAR	&	12.23	&	4.36	&	0.54	&	0.09	&	0.99	&	0.03	\\
	        &	MAR	&	12.81	&	4.04	&	0.55	&	0.07	&	0.98	&	0.04	\\
$\;\;$MI-RWC	&	SMAR	&	7.44	&	5.17	&	0.40	&	0.16	&	0.95	&	0.05	\\
	        &	MAR	&	8.40	&	5.72	&	0.43	&	0.15	&	0.95	&	0.06	\\
$\;\;$MI-seq	&	SMAR	&	2.41	&	1.88	&	0.19	&	0.12	&	0.91	&	0.08	\\
	          &	MAR	&	2.99	&	2.28	&	0.23	&	0.12	&	0.91	&	0.07	\\
{\em Selection prop.: $\geq 0.9$} 	&		&		&	     	&		&		&		&		\\
$\;\;$MI-lasso	&	SMAR	&	7.46	&	3.52	&	0.41	&	0.12	&	0.98	&	0.04	\\
	        &	MAR	&	8.21	&	3.35	&	0.44	&	0.10	&	0.97	&	0.05	\\
$\;\;$MI-RWC	&	SMAR	&	4.35	&	3.67	&	0.28	&	0.16	&	0.93	&	0.07	\\
	        &	MAR	&	5.14	&	4.15	&	0.31	&	0.16	&	0.93	&	0.07	\\
$\;\;$MI-seq	&	SMAR	&	1.31	&	1.24	&	0.12	&	0.10	&	0.88	&	0.08	\\
	        &	MAR	&	1.67	&	1.58	&	0.14	&	0.11	&	0.88	&	0.08	\\
{\em  Selection prop.: $1$} 	&		&		&		&		&		&		&		      \\
$\;\;$MI-lasso	&	SMAR	&	3.87	&	2.20	&	0.27	&	0.12	&	0.96	&	0.06	\\
	        &	MAR	&	4.43	&	2.39	&	0.30	&	0.11	&	0.95	&	0.06	\\
$\;\;$MI-RWC	&	SMAR	&	2.04	&	2.00	&	0.17	&	0.13	&	0.88	&	0.09	\\
	        &	MAR	&	2.39	&	2.43	&	0.18	&	0.14	&	0.89	&	0.08	\\
$\;\;$MI-seq	&	SMAR	&	0.53	&	0.69	&	0.05	&	0.07	&	0.83	&	0.10	\\
	        &	MAR	&	0.70	&	0.86	&	0.07	&	0.08	&	0.83	&	0.10	\\
\botrule
\end{tabular}
\end{table}

The reference values for evaluating the performance of the MI-based approaches are those achieved by XCDW: an FDR of 0.17 under SMAR and 0.20 under MAR, a TPR of 0.91 under SMAR and 0.90 under MAR. Notably, although the XCDW approach is valid only under SMAR (i.e., missingness depending only on non-null variables), its performance is minimally affected when applied to MAR data. For all approaches, switching from SMAR to the more general MAR entails a modest increase in FDR, while the TPR remains essentially unchanged.

Focusing on MI-based knockoff approaches, MI-RWC demonstrates a performance comparable to that of XCDW only when variables are retained if selected in all imputed datasets (i.e., selection proportion equal to 1). Under this condition, MI-RWC achieves a Monte Carlo mean FDR of 0.17–0.18 and a mean PFER close to 2, albeit with a slight reduction in TPR (0.88–0.89). Differently, with a selection proportion $\geq 0.9$, the FDR rises to approximately 0.30 and the PFER increases to over 4, with both metrics worsening further at smaller selection proportions.  

A markedly different scenario emerges when evaluating the performance of MI-seq. Specifically, when variables are retained if selected in at least 80\% of the imputed datasets, the FDR and TPR are very close to those achieved by XCDW. Increasing the selection proportion further reduces the FDR but comes at the cost of a slight decrease in TPR, which declines from 0.91 (for a selection proportion $\geq 0.8$) to 0.88 (if $\geq 0.9$), reaching a minimum of 0.83 when variables are retained only if selected in all imputed datasets.

\subsection{Results of Monte Carlo simulations for varying rates of missing values}\label{sec:ressim2}

We extend the simulation study presented in the previous section to investigate the performance of MI-based approaches under varying rates of missing data. First of all, we check for the robustness of the proposed methods in the absence of missing values (i.e., with complete data). In this setting, and with a nominal PFER set to 2, both MI-RWC and MI-seq effectively control the PFER, yielding empirical values of 1.37 and 1.51, respectively (FDR: 0.11 and 0.12; TPR: 0.99 and 1.00).

Table \ref{tab:sim2} reports the values of $\widehat{PFER}$, $\widehat{FDR}$, and $\widehat{TPR}$ for increasing rates of missing values: 10\%, 32\%, and 45\%, generated according to eq. \eqref{eq:miss}.

\begin{table}[!ht]
\caption{Variable selection with XCDW, MI-lasso, MI-RWC, and MI-seq, for increasing rates of missing values (about 10\%, 32\%, and 45\%), by selection proportion: Monte Carlo means of PFER, FDR, and TPR.}
\label{tab:sim2}%
\begin{tabular}{@{}llrrrrrrrrrrr@{}}
\toprule
Method	&		\multicolumn{4}{c}{PFER}			&	\multicolumn{4}{c}{FDR}	&			\multicolumn{4}{c}{TPR}			\\				
	&	10\%	&	    	&	32\%	&	45\%	&	10\%	&	    	&	32\%	&	45\%	&	10\%	&	    	&	32\%	&	45\%	\\
    \midrule
XCDW	&	0.00	&		&	2.51	&	56.53	&	0.00	&		&	0.20	&	0.86	&	0.905	&		&	0.995	&	0.93	\\
{\em Selection prop.: $\geq 0.60$} 	&		&		&		&		&		&		&		&		&		&		&		&		\\
	&		&		&		&		&		&		&		&		&		&		&		&		\\
MI-lasso	&	14.07	&		&	24.37	&	80.58	&	0.58	&		&	0.70	&	0.89	&	1.00	&		&	0.99	&	1.00	\\
	&		&		&		&		&		&		&		&		&		&		&		&		\\
MI-RWC	&	2.09	&		&	18.06	&	79.56	&	0.16	&		&	0.61	&	0.89	&	0.99	&		&	0.98	&	0.98	\\
	&		&		&		&		&		&		&		&		&		&		&		&		\\
MI-seq	&	2.18	&		&	6.76	&	12.62	&	0.17	&		&	0.40	&	0.55	&	0.99	&		&	0.95	&	0.82	\\
Selection prop.: $\geq 0.70$ 	&		&		&		&		&		&		&		&		&		&		&		&		\\
	&		&		&		&		&		&		&		&		&		&		&		&		\\
MI-lasso	&	11.60	&		&	17.91	&	70.96	&	0.53	&		&	0.63	&	0.87	&	1.00	&		&	0.99	&	1.00	\\
	&		&		&		&		&		&		&		&		&		&		&		&		\\
MI-RWC	&	1.61	&		&	12.56	&	68.55	&	0.13	&		&	0.52	&	0.88	&	0.99	&		&	0.97	&	0.95	\\
	&		&		&		&		&		&		&		&		&		&		&		&		\\
MI-seq	&	1.65	&		&	4.58	&	7.44	&	0.13	&		&	0.31	&	0.44	&	0.99	&		&	0.93	&	0.77	\\
Selection prop.: $\geq 0.80$ 	&		&		&		&		&		&		&		&		&		&		&		&		\\
	&		&		&		&		&		&		&		&		&		&		&		&		\\
MI-lasso	&	9.41	&		&	12.81	&	56.25	&	0.48	&		&	0.55	&	0.83	&	1.00	&		&	0.98	&	0.99	\\
	&		&		&		&		&		&		&		&		&		&		&		&		\\
MI-RWC	&	1.19	&		&	8.40	&	51.46	&	0.10	&		&	0.43	&	0.84	&	0.98	&		&	0.95	&	0.90	\\
	&		&		&		&		&		&		&		&		&		&		&		&		\\
MI-seq	&	1.21	&		&	2.99	&	4.04	&	0.10	&		&	0.23	&	0.30	&	0.98	&		&	0.91	&	0.72	\\
Selection prop.: $\geq 0.90$ 	&		&		&		&		&		&		&		&		&		&		&		&		\\
	&		&		&		&		&		&		&		&		&		&		&		&		\\
MI-lasso	&	7.03	&		&	8.21	&	37.39	&	0.40	&		&	0.44	&	0.74	&	1.00	&		&	0.97	&	0.97	\\
	&		&		&		&		&		&		&		&		&		&		&		&		\\
MI-RWC	&	0.82	&		&	5.14	&	31.81	&	0.07	&		&	0.31	&	0.77	&	0.98	&		&	0.93	&	0.82	\\
	&		&		&		&		&		&		&		&		&		&		&		&		\\
MI-seq	&	0.84	&		&	1.67	&	2.00	&	0.07	&		&	0.14	&	0.19	&	0.98	&		&	0.88	&	0.65	\\
Selection prop.: $1.00$ 	&		&		&		&		&		&		&		&		&		&		&		&		\\
	&		&		&		&		&		&		&		&		&		&		&		&		\\
MI-lasso	&	4.91	&		&	4.43	&	16.38	&	0.32	&		&	0.30	&	0.56	&	1.00	&		&	0.95	&	0.86	\\
	&		&		&		&		&		&		&		&		&		&		&		&		\\
MI-RWC	&	0.48	&		&	2.39	&	12.11	&	0.04	&		&	0.18	&	0.60	&	0.97	&		&	0.89	&	0.63	\\
	&		&		&		&		&		&		&		&		&		&		&		&		\\
MI-seq	&	0.47	&		&	0.70	&	0.59	&	0.04	&		&	0.07	&	0.08	&	0.97	&		&	0.83	&	0.53	\\
\botrule
\end{tabular}
\end{table}

The simulation results highlight several key patterns. 
Overall, as expected the performance of all compared methods deteriorates as the proportion of missing data increases. 
When the amount of missing data is low, the XCDW performs exceptionally well, while the two MI-based knockoffs  approaches yield comparable results. 

However, when missing values increase,  a clear performance gap emerges between MI-RWC and MI-seq, with the latter showing a noticeable advantage. For instance, at about 32\% of missing values, MI-RWC  never controls for the PFER, even when the selection proportion is equal to 1 (empirical PFER settles at 2.39). Differently, the MI-seq   reports an empirical PFER of 2 with a selection proportion equal to 0.90 and falls below  2 (i.e., 0.70) with a selection proportion of 1. 
In this scenario, the difference between XCDW and MI-seq  also tends to diminish, as already discussed in Section \ref{sec:ressim1}.

At the highest level of missing data considered (45\%), the only method that still manages to control the PFER is the MI-seq - provided a high selection threshold (e.g., 90\%) is used - although this comes at the cost of a considerable drop in TPR, reaching a value of 0.53. On the other hand, with such rates of missing values  XCDW performs definitely worse than the MI-based knockoffs procedures, with an empirical PFER equal to 56.53.

These findings suggest that the optimal selection proportion to adopt in the MI-based knockoff procedures is strongly influenced by the extent of missing data: the higher the proportion of missing values, the more stringent the selection rule must be. In general, the commonly recommended majority rule for multiple imputation  \citep[as suggested, for instance, by ][]{wood2008} appears insufficient in our setting. Namely, in our simulation framework, the majority rule corresponds to a selection proportion of $\geq 0.60$, which does not adequately control the error rate under medium-high levels of missingness.

\subsection{Remarks on simulation results}
In summary, the MI-seq approach represents a valuable alternative to the XCDW method. Based on the simulation results described above, there is evidence that the  selection proportion depends on the rate of missing values. For a low missing rate (approximately 10\%), retaining a variable if it is selected in at least 70\% of the imputed datasets is sufficient to control the PFER while maintaining high TPR values. When the missing rate increases to 32\%, this threshold rises to 90\%. For very high missing rates (around 45\%), the general recommendation is to retain variables selected in at least 90\% of the imputed datasets; however, this comes at the cost of a substantial reduction in TPR. 

It is worth noting that the simulation design used here is the one proposed by \cite{xie:2023} to validate their method. In particular, the continuous variables are assumed to be normally distributed, which is important for the XCDW method that is based on the likelihood. Indeed,  deviations from normality could negatively affect the XCDW approach. On the other hand, the normality assumption is irrelevant for the MI-based approaches, since the imputation phase for continuous variables exploits least-squares linear models. Moreover, the simulation study considers only binary and continuous variables, neglecting ordered and unordered categorical variables. Ordinal variables are possible in XCDW, even if the selection procedure should be adapted to properly account for the ordering of the categories. On the other hand, unordered categorical variables  are not handled by XCDW, which is a limitation in some applications. Indeed, the analysis of INVALSI data on math test scores presented in the next section requires the ability of the sequential knockoff filter to deal with   categorical variables.

\section{Variable selection via Knockoffs for INVALSI data on mathematics test scores}
\label{sec:analysis}

The INVALSI is the national institute for the evaluation of the Italian school system. Specifically,  it performs the evaluation of student achievements in Mathematics, Science, Reading and English every year for different grades. The evaluation is based on multiple-choice questions. The selection of the set of items relies on internationally validated methods based on the Rasch model \citep{rasch:1960},  similarly, for example, to TIMSS \citep{vonDavier:2024}. The evaluation test is administered to all students in all schools, to assess the performance of the schools. An additional background questionnaire is administered to the students of a sample of schools to study the relationship between achievement and students' background characteristics. We focus on data from the INVALSI sample survey of the year 2022-23 on pupils in grade 5, corresponding to the last year of primary school. The dataset includes 16,828 students nested in 501 schools.

Specifically, we consider the achievement in mathematics, which is a latent ability measured by a set of dichotomously scored items (1: correct answer; 0: wrong answer). 
INVALSI provides the response pattern to the items and a Rasch-based estimate of the ability. 
The average sample score in mathematics is 191.82 points, with a standard deviation of 41.07.

The data set includes 30 student-level variables of different types:
\begin{description}
    \item[\hspace{4ex} -] binary  (e.g., gender, availability of a personal computer);
    \item[\hspace{4ex} -] ordinal (e.g. mother's education);
    \item[\hspace{4ex} -] unordered categorical (e.g. mother’s occupation).
 \end{description}

The data set also includes the geographical area of the school, which does not enter the variable selection process, as it is used as a stratification variable.

To study the relationship between the background characteristics and pupils' achievement, while accounting for their clustering into schools, we specify a random intercept model \citep{snijders:2011}. Denoting the pupils with $i$ (level 1) and the schools with $s$ (level 2), the model is specified as: 

\begin{equation}
\label{eq:model}
	y_{is}=\mathbf{x}_{is} \bm{\beta}  + \mathbf{z}_{s}\bm{\gamma} + u_s + e_{is}
 \vspace{1em}
\end{equation}
where $y_{is}$ is the mathematics score of pupil $i$ of school $s$, $\mathbf{x}_{is}$ is the row vector of level 1 predictors with coefficients $\bm{\beta}$ (including the intercept) and $\mathbf{z}_{s}$ is the row vector of level 2 predictors with coefficients $\bm{\gamma}$.
The level 1 errors are assumed to be independent and identically distributed, $e_{is} \sim N(0, \sigma_e^2)$, and the level 2 errors (random effects) are assumed to be independent and identically distributed,  $u_s \sim N(0, \sigma_u^2)$. Moreover, the errors at the two levels are assumed to be independent.

The background variables are highly correlated, and there are no theoretical guidelines to choose among them, calling for a data-driven selection procedure.

Moreover, a critical issue of the data is the relevant rate of missing values for many student-level variables, ranging from $1.6$\% to $25.5$\%.
Specifically, the variables with the largest rates of missing values are Father's and Mother's education and occupation (from 23.4\% to 25.5\%), followed by Italian language at home, Dialect spoken at home, Italian used with friends, and Nursery school (about 10\%).
A complete-case analysis would halve the number of students with a large loss of power and, especially, a potentially high bias given that the missing mechanism cannot be assumed to be missing completely at random. 
Rather, a MAR mechanism with missingness depending on observed values may be plausible, so multiple imputation is a viable approach to preserve the original sample size and avoid bias in the estimators.

To perform variable selection while handling the missing data issue, we follow our proposed procedure described in Section \ref{sec:methodproposal}, 
that is, the MI-seq. It is worth noting that the other procedures (i.e., XCDW and MI-RWC) are not suitable, as they do not handle variables with unordered categories, such as Immigrant status, Father's occupation, and Mother's occupation.

Specifically, we implement MICE with 10 imputed datasets using the \texttt{mice} package of \texttt{R} \citep{buuren:2011}. Each variable is imputed with a model suitable for its measurement scale: 
logit model for binary variables,  cumulative logit model for ordinal variables and multinomial logit model for unordered categorical variables.  

For each imputed dataset, the predictors are selected using the 
 sequential knockoffs procedure of  \cite{kormaksson:2021},  as implemented in the  \texttt{R} package  \texttt{knockofftools} \citep{zimmermann:2024}  with options PFER=2 and \texttt{method="sparseseq"}. 
Unordered categorical and ordinal variables are coded using a set of dummy variables with a baseline category.
In the knockoffs filter, we use group lasso \citep{yuan2006} instead of standard lasso, to avoid the selection depending on the arbitrary choice of the baseline category. In this way, for each predictor, the corresponding set of dummy variables is jointly selected or discarded. 
For each imputed dataset, the procedure runs 31 times and a predictor is retained if selected in at least 50\% of the runs. 

Table \ref{tab:selection} shows the predictors with the proportion selected over the 10 imputed datasets by the sequential knockoffs procedure. According to the findings of the simulation study illustrated in Section \ref{sec:simulation}, a predictor is included in the final model if selected in at least 8 out of 10 imputed datasets: this is denoted by a check-mark in Table  \ref{tab:selection}. 
Overall, 22 out of 30 student-level variables have been selected.

\begin{table}[!ht]
\caption{Variable selection via sequential knockoff: proportion selected over 10 imputed datasets and retained status ($\checkmark$ if proportion $\ge 0.8$). INVALSI grade 5 sample survey, year 2022-23.}
\label{tab:selection}
\begin{tabular}{llrc}
\hline
    \textit{Variable}	&	\textit{Levels}	&	\textit{Pr.}	& \textit{Ret.}\\ \hline
	Male	&	1=yes, 0=no	&	1	&$\checkmark $\\
	\# brothers	&	ordinal, 5 levels	&	1	&$\checkmark $\\
	\# sisters	&	ordinal, 5 levels	&	1	&$\checkmark $\\
	At home, you have:	&		&	&	\\
\hspace{0,5ex}	     quiet space	&	1=yes, 0=no	&	0.9	&$\checkmark $\\
\hspace{0,5ex}	     computer	&	1=yes, 0=no	&	1	&$\checkmark $\\
\hspace{0,5ex}	     personal desk	&	1=yes, 0=no	&	0.3	&\\
\hspace{0,5ex}	     software	&	1=yes, 0=no	&	0	&\\
\hspace{0,5ex}	     internet connection	&	1=yes, 0=no	&	1	&$\checkmark $\\
\hspace{0,5ex}	     personal room	&	1=yes, 0=no	&	1	&$\checkmark $\\
\hspace{0,5ex}	     classical books	&	1=yes, 0=no	&	0	&\\
\hspace{0,5ex}	     artworks	&	1=yes, 0=no	&	0	&\\
\hspace{0,5ex}	      technical manuals	&	1=yes, 0=no	&	0	&\\
\hspace{0,5ex}	      dictionary	&	1=yes, 0=no	&	1 &$\checkmark $	\\
\hspace{0,5ex}	      personal tablet	&	1=yes, 0=no	&	1	&$\checkmark $\\
\hspace{0,5ex}	      smartphone	&	1=yes, 0=no	&	1	&$\checkmark $\\
\hspace{0,5ex}	\#  books	&	ordinal, 5 levels	&	1	&$\checkmark $\\
	Student origin and language	&		&		&\\
\hspace{0,5ex}	Student born in Italy	&	1=yes, 0=no	&	1	&$\checkmark $\\
\hspace{0,5ex}	Immigrant staus	&	1=native; 2=1st gen immigrant; 3=2nd gen imm.	&	0 &	\\
\hspace{0,5ex}	Italian language at home	&	1=yes, 0=no	&	1	& $\checkmark $\\
\hspace{0,5ex}	Dialect spoken at home	&	1=yes, 0=no	&	1	&$\checkmark $\\
\hspace{0,5ex}	Italian used with friends	&	1=yes, 0=no	&	1	&$\checkmark $\\
	Parents	&		&		&\\
\hspace{0,5ex}	Father born in Italy	&	1=yes, 0=no	&	0.2	&\\
\hspace{0,5ex}	Mother born in Italy	&	1=yes, 0=no	&	1	&$\checkmark $\\
\hspace{0,5ex}	Father’s education	&	1=compulsory, 2=high sch., 3=college, 4=master+	&	1	&$\checkmark $\\
\hspace{0,5ex}	Mother’s education	&	1=compulsory, 2=high sch., 3=college, 4=master+    &	1	&$\checkmark $\\
\hspace{0,5ex}	Father’s occupation	&	1=unenmployed 2=manager, clerk 3=self-employed  	&	1	&$\checkmark $\\
		&	\hspace{1ex} 4=blue collar 5=professional 6=housewife, retired 	&		&\\
\hspace{0,5ex}	Mother’s occupation	&	 1=unenmployed 2=manager, clerk 3=self-employed &	1	&$\checkmark $\\
		&	\hspace{1ex} 4=blue collar 5=professional 6=housewife, retired 	&	&	\\
	Student career	&		&		&\\
\hspace{0,5ex}	Nursery school	&	(for pupils aged 3-5 years) 1=yes, 0=no	&	1	&$\checkmark $\\
\hspace{0,5ex}	Regular student	&	1=yes, 0=no	&	1	&$\checkmark $\\
\hspace{0,5ex}	School weekly hours$>$30	&	1=yes, 0=no	&	0	&\\
School area	&	1=NW; 2=NE; 3=Center; 4= South; 5= S-Islands	&	\multicolumn{2}{c}{\em retained by default}\\
\hline
\end{tabular}
\end{table}

The sequential knockoff procedure gives parameter estimates based on the group lasso for a standard linear model, i.e. without random effects. 
This is not a limitation for selecting the predictors, since in the linear model, there is an equivalence between the regression coefficients conditional on the random effects and the marginal ones \citep{ritz2004}.
Notwithstanding, we need to fit a random effect model with the selected variables for two reasons: (i) to obtain standard errors properly accounting for the school clustering; (ii) to avoid the downward bias implicit in lasso-based procedures \citep[e.g.][]{belloni:2013}. For these reasons, we fit the random intercept model (\ref{eq:model}) with the selected predictors to each imputed dataset via maximum likelihood using the \texttt{R} package 
\texttt{lme4} \citep{lme4}.
The results across the 10 imputed datasets are combined using the Rubin's rules to obtain point estimates and standard errors.

Table \ref{tab:variances} reports the residual variances and the Intraclass Correlation Coefficient (ICC) for the model without predictors (null model) and for the model with the predictors selected by sequential knockoffs (full model; see Table \ref{tab:selection}). The null model shows that the variability in the mathematics scores due to school clustering is $13.8\%$. The predictors explain about $23\%$ of the school-level variance and about $15\%$ of the pupil-level variance. Given the included predictors, the residual ICC in the final model is $12.6\%$, indicating that the school clustering is still relevant.

Table \ref{tab:results} presents the point estimates and corresponding standard errors for the full model, obtained through  Rubin's rules. It is worth noting that the standard errors account for the variability due to multiple imputation; however, they ignore the uncertainty implicit in the variable selection process through knockoffs and, thus, are likely to be underestimated, raising a problem for post-selection inference. Anyway, we are mainly interested in the effects of the selected predictors, whose direction (i.e., sign of the regression coefficients) is well detected by knockoffs methods \citep{barber:candes:2019}, thus in the following we limit the comments to the point estimates. 

In general, the effects align with expectations, with a few notable exceptions. Interestingly, ownership of a personal tablet or smartphone is associated with lower mathematics scores. Since the test evaluates mathematics proficiency acquired during primary school, this result could reflect the potentially harmful impact of device usage by children \citep{BELAND201661}. Additionally, owning such devices may signal a family environment less conducive to learning. 

Mother's and father's education have similar effects, monotonically increasing from compulsory to master or more. On the other hand, the effects of occupation are much greater for the father. The largest effect pertains to the number of books at home, which is a proxy of the cultural environment.


\begin{table}[!ht]
\caption{Random intercept null model and full model with selected predictors (Table \ref{tab:selection}):  combined estimates of the residual variances based on 10 imputed datasets. INVALSI grade 5 sample survey, year 2022-23.}
\label{tab:variances}
\begin{tabular}{lrrr}
\hline
Residual variances	&	null model	&	full model 	&	\% explained	\\ \hline
Level 2 (school)	&	233.30	&	178.94	&	23.30	\\
Level 1 (pupil)	&	1,461.94	&	1,239.97	&	15.18	\\
Total	&	1,695.24	&	1,418.91	&	16.30	\\ \hline
ICC	&	13.76\% &	12.61\% &		\\ \hline
\end{tabular}
\end{table}

\begin{table}[!ht]
\caption{Random intercept model with selected predictors: combined estimates of the regression coefficients and standard errors based on 10 imputed datasets. INVALSI grade 5 sample survey, year 2022-23.}
\label{tab:results}
\begin{tabular}{lrrr}
\hline
	Variable	&	Estimate	&	Std. Error	
    \\ \hline
	Intercept	&	124.26	&	4.016	\\
	Male	&	9.55	&	0.568	\\
	\# brothers (ref. 0)	&		&		\\
\hspace{2 ex}	1	&	-4.33	&	0.652	\\
\hspace{2 ex}	2	&	-6.98	&	1.082	\\
\hspace{2 ex}	3	&	-11.92	&	1.961	\\
\hspace{2 ex}	$\ge$ 4	&	-12.70	&	2.929	\\
	\# sisters (ref. 0)	&		&		\\
\hspace{2 ex}	1	&	-3.00	&	0.645	\\
\hspace{2 ex}	2	&	-5.38	&	1.109	\\
\hspace{2 ex}	3	&	-9.55	&	2.472	\\
\hspace{2 ex}	$\ge$ 4	&	-11.28	&	3.776	\\
	At home, you have:	&		&		\\
\hspace{0.5ex}	quiet space	&	1.20	&	0.832	\\
\hspace{0.5ex}	computer	&	3.39	&	0.612	\\
\hspace{0.5ex}	internet connection	&	4.49	&	0.862	\\
\hspace{0.5ex}	personal room	&	-2.58	&	0.613	\\
\hspace{0.5ex}	dictionary	&	8.47	&	1.003	\\
\hspace{0.5ex}	personal tablet	&	-4.11	&	0.586	\\
\hspace{0.5ex}	smartphone	&	-4.29	&	0.661	\\
	\# of books (ref. 0-10)	&		&		\\
\hspace{2 ex}	11-25	&	3.96	&	0.933	\\
\hspace{2 ex}	26-100	&	12.34	&	0.966	\\
\hspace{2 ex}	101-200	&	15.39	&	1.126	\\
\hspace{2 ex}	$>$ 200	&	17.33	&	1.262	\\
	Student origin and language:	&		&		\\
\hspace{0.5ex}	Student born in Italy	&	6.31	&	1.673	\\
\hspace{0.5ex}	Italian language at home	&	4.56	&	1.071	\\
\hspace{0.5ex}	Dialect spoken at home	&	-1.91	&	0.617	\\
\hspace{0.5ex}	Italian used with friends	&	9.87	&	1.540	\\
\hspace{0.5ex}	Mother born in Italy	&	-1.69	&	0.975	\\
	Father’s education (ref. compulsory)	&		&		\\
\hspace{2 ex}	high school	&	5.31	&	0.921	\\
\hspace{2 ex}	college	&	8.72	&	1.541	\\
\hspace{2 ex}	master or more	&	10.49	&	1.333	\\
	Mother’s education (ref. compulsory)	&		&		\\
\hspace{2 ex}	high school	&	5.54	&	0.885	\\
\hspace{2 ex}	college	&	6.75	&	1.342	\\
\hspace{2 ex}	master or more	&	10.36	&	1.316	\\
	Father’s occupation (ref. unempl.)	&		&		\\
\hspace{2 ex}	manager, clerk	&	7.11	&	1.857	\\
\hspace{2 ex}	 self-employed	&	5.60	&	1.858	\\
\hspace{2 ex}	blue collar	&	2.90	&	1.828	\\
\hspace{2 ex}	professional	&	5.58	&	1.922	\\
\hspace{2 ex}	housewife, retired	&	2.68	&	3.546	\\
	Mother’s occupation (ref. unemployed)	&		&		\\
\hspace{2 ex}	manager, clerk	&	2.11	&	1.483	\\
\hspace{2 ex}	 self-employed	&	1.32	&	1.659	\\
\hspace{2 ex}	blue collar	&	-0.76	&	1.522	\\
\hspace{2 ex}	professional	&	2.80	&	1.637	\\
\hspace{2 ex}	housewife, retired	&	-0.62	&	1.495	\\
	Student career:	&		&		&		\\
\hspace{0.5ex}	Nursery school	&	8.90	&	1.520	\\
\hspace{0.5ex}	Regular student	&	5.18	&	2.287	\\
	School area (ref. North-West)	&		&		\\
\hspace{2 ex}	North-East	&	2.96	&	2.147	\\
\hspace{2 ex}	Center	&	5.59	&	2.161	\\
\hspace{2 ex}	South	&	8.65	&	2.089	\\
\hspace{2 ex}	South-Islands	&	-1.39	&	2.151	\\
\hline
\end{tabular}
\end{table}

\newpage

\section{Final remarks}
\label{sec:final}
In this contribution, we proposed an approach, referred to as MI-seq, for variable selection in the presence of missing values using knockoffs. The proposed method consists of the following sequential phases: first, missing values are addressed through multiple imputation techniques; second, for each imputed dataset, variables are selected using a knockoffs method, and only those variables selected in at least a predefined minimal proportion of datasets are retained; finally, a suitable regression model is estimated based on the previously selected variables. This multi-step structure provides substantial flexibility, distinguishing the MI-seq from the recent contribution by \cite{xie:2023}, referred to here as XCDW, which, to the best of our knowledge, is the only one that applies the knockoff method in the presence of missing data. Furthermore, MI-seq does not require any assumption on the distribution of the continuous variables and permits handling unordered categorical variables in addition to continuous, binary and ordinal. Finally, MI-seq yields results valid using the standard Missing At Random (MAR) assumption instead of the more restrictive stronger MAR assumption of XCWD, where the missing data mechanism depends only on non-null variables.

These advantages of MI-seq are made possible because of the following features: (i) the use of multiple imputation to deal with missing values, allowing to deal with any type of variables under the MAR assumption, and (ii) the selection of variables based on the sequential knockoffs procedure of \cite{kormaksson:2021} and \cite{zimmermann:2024} that allows to generate knockoff copies of unordered categorical variables based on the multinomial logit model, differently from standard knockoffs procedures \citep[e.g., see ][]{ren:2023}, which are designed for continuous variables. 
Moreover, separating the imputation and modeling steps allows to specify any kind of model, such as the multilevel model adopted for the analysis of the INVALSI data on mathematics scores. 

To evaluate the performance of the MI-seq we conducted a simulation study under the scenario proposed by \cite{xie:2023}, obtaining similar performances. Interestingly, MI-seq performed better than the approach based on integrating the multiple imputation with the derandomized knockoff procedure of \cite{ren:2023} (referred to in the text as MI-RWC), likely because in the simulation half of the variables are binary. These results proved the validity of the proposed MI-seq method, which we developed to deal with more complex scenarios such as the one involved in the analysis of INVALSI data with several unordered categorical variables and a hierarchical structure calling for a multilevel model.

A limitation of the proposed MI-seq approach is that it does not strictly control the PFER or the FDR, contrary to knockoff filters for complete data and the XCDW method of \cite{xie:2023} for SMAR data. In fact, any MI-based procedure requires as a further tuning parameter the minimal proportion of imputed datasets for ultimately retaining a variable and the PFER is inversely related to this proportion. In the simulation study, the optimal proportion was $0.8$: even if this finding cannot be generalized, a minimal proportion of $0.8$ can be adopted as the default. The choice of this 
proportion is not always critical since, in applications, it may happen that the selection is not very sensitive to this parameter. For example, in our case study, the non-null variables were clearly identified since the actual proportions were either $\ge 0.9$ or $\le 0.3$.

A further challenging issue concerns the post-selection inference in the presence of missing data. Indeed, the knockoff methods are designed to control the FDR (or PFER) and they can even control the proportion of wrong signs in the regression coefficients \citep{barber:candes:2019}. However, the standard errors obtained by fitting the model with selected predictors are underestimated, as they ignore the uncertainty of the selection process. A general solution to this issue consists in splitting the   data  into two sets, one for the variable selection and the other  for the model fitting and inference \citep[see, for instance, ][]{WassermanRoeder2009, GarciaRasines:2023}. However, the implementation of this approach is not straightforward when multiple imputation is adopted. For instance, the data splitting can be done before or after the multiple imputation of missing values. This is an interesting topic for further research. 

Finally, a general issue about knockoffs, not related to missing data, is their performance in selecting cluster-level predictors in multilevel analysis. In fact, the available methods to generate the knockoffs do not account for the levels of the predictors, so a copy of a cluster-level variable is produced as if it were an individual-level variable, ignoring the constraint that it has to be constant within the clusters. 
This limitation does not apply in the INVALSI case study because all predictors are at the student level (with the exception of the geographical area, which is not included in the selection procedure).
However, applications of multilevel models often involve many cluster-level predictors; thus, future research should investigate whether the selection process of cluster-level predictors is negatively affected by the use of knockoffs designed for individual-level predictors and, possibly, to devise effective solutions.

\bibliography{biblio_knockoffs}

\end{document}